\documentclass[a4paper,fleqn]{cas-sc}

\usepackage[numbers]{natbib}

\usepackage{booktabs}
\usepackage{graphicx}
\usepackage{caption}
\usepackage{subcaption}
\usepackage{placeins}

\let\printorcid\relax 

\def\tsc#1{\csdef{#1}{\textsc{\lowercase{#1}}\xspace}}
\tsc{WGM}
\tsc{QE}
\tsc{EP}
\tsc{PMS}
\tsc{BEC}
\tsc{DE}

\begin{document}
\let\WriteBookmarks\relax
\def\floatpagepagefraction{1}
\def\textpagefraction{.001}

\shorttitle{New Monitoring Interface for the AMS Experiment}

\shortauthors{R. K. Hashmani, M. Konyushikhin, B. S. Shan et~al.}

\title [mode = title]{New Monitoring Interface for the AMS Experiment}                      



%
\author[1]{Raheem Karim Hashmani}

\cormark[1]


\ead{raheem.hashmani@cern.ch}

\affiliation[1]{organization={Middle East Technical University (METU)},
    city={Ankara},
    citysep={}, 
    postcode={06800}, 
    country={Turkey}}

\author[2]{Maxim Konyushikhin}

\affiliation[2]{organization={Massachusetts Institute of Technology (MIT)},
    city={Cambridge},
    state={Massachusetts},
    statesep={}, 
    postcode={02139},
    country={USA}}

\author[3]{Baosong Shan}
\ead{baosong.shan@cern.ch}

\affiliation[3]{organization={Beihang University (BUAA)},
    city={Beijing},
    postcode={100191},
    country={China}}

\author[2]{Xudong Cai}

\author[1]{Melahat Bilge Demirköz}

\cortext[cor1]{Corresponding author}



\begin{abstract}
The Alpha Magnetic Spectrometer (AMS) is constantly exposed to harsh condition on the ISS. As such, there is a need to constantly monitor and perform adjustments to ensure the AMS operates safely and efficiently. With the addition of the Upgraded Tracker Thermal Pump System, the legacy monitoring interface was no longer suitable for use. This paper describes the new AMS Monitoring Interface (AMI). The AMI is built with state-of-the-art time series database and analytics software. It uses a custom feeder program to process AMS Raw Data as time series data points, feeds them into InfluxDB databases, and uses Grafana as a visualization tool. It follows modern design principles, allowing client CPUs to handle the processing work, distributed creation of AMI dashboards, and up-to-date security protocols. In addition, it offers a more simple way of modifying the AMI and allows the use of APIs to automate backup and synchronization. The new AMI has been in use since January 2020 and was a crucial component in remote shift taking during the COVID-19 pandemic. 
\end{abstract}




\maketitle

\section{Introduction}

The Alpha Magnetic Spectrometer (AMS) is a general-purpose high-energy particle physics detector. It was installed on the International Space Station (ISS) on 19 May 2011 to conduct a unique, long-duration mission of fundamental physics research in space. Its main objectives include searching for antimatter, investigating dark matter, and analyzing cosmic rays using the 6 detectors onboard. These detectors include the gaseous Xe/CO2 Transition Radiation Detector (TRD), the nine layer Silicon Tracker, the lead/scintillator fiber Electromagnetic Calorimeter (ECAL), two Time of Flight Detectors (ToF), and the aerogel/NaF Ring Imaging Cherenkov (RICH). At the very center is a permanent Nd-Fe-B magnet which generates a magnetic field of 0.15 T \cite{Ting2013}. AMS has been continuously collecting cosmic ray particles since its installation onboard the ISS and has collected over 200 billion cosmic ray events.

Due to the extreme conditions in space, a myriad of sensors were installed onboard the AMS that measure the temperature, pressure, current, voltage, and other health data of its various components. This monitoring data is transmitted together with the cosmic ray data to the ground with an average transmission speed of 10 MBit/sec \cite{Choutko2015ComputingExperiment}.

The detector and its components are continuously exposed to space radiation, extreme temperature variations, solar effects, and changes in the ISS orientation. One of the most challenging areas of the AMS operations is the continuous real-time monitoring and adjustment of temperatures, voltages, currents, and pressures. Due to the ISS orbit lasting \textasciitilde 93 minutes, these values change drastically, going from maximal to minimal and back again within that time span. Certain measurements, such as temperature, can change from -60° C to +80° C within a single orbit without the use of controlled heaters and thermal blankets. As such, real-time monitoring is required to prevent certain values from going beyond their range of operational values.

The legacy AMS Monitoring Interface (AMI) is a set of software designed to allow quick access to the health status of the AMS and its various components. Experts at the CERN Payload Operations and Control Center (POCC), at CERN, Geneva, and the AMS Asia POCC in Taiwan use this interface to monitor the AMS 24 hours a day.

In recent years, the performance of the legacy AMI had degraded and become inefficient, taking upwards of 45 seconds to show certain complex plots. The design philosophy had become outdated, generating static plots from the backend and pushing them to the frontend website. It used only one of the incoming data streams from the ISS at any given time and was therefore not taking advantage of the redundancy measures of the AMS's two main streams: real-time and buffered. It was built with a heavy reliance on scripting and required changes be done by specific people who knew how to rework the scripts. Lastly, it had to be supplemented with various other custom-built monitoring console software to provide additional capabilities. Discussion to replace it with a modern, up-to-date monitoring system had started to take place since launch.

The final push came in 2019. In the latter half of 2014, the old pumps on the mechanically pumped two-phase CO2 cooling loop \cite{delil2002development}, which circulated coolant to the Silicon Tracker, reached their designed lifetime. In order to prolong the lifetime of the AMS experiment, an upgraded cooling system known as the Upgraded Tracker Thermal Pump System (UTTPS) was designed, manufactured, tested, space qualified, and deployed to the ISS in November 2019. This upgrade brought along additional sensors to be installed onboard the AMS, increasing the number of data sources.

Due to the design philosophy of the legacy AMI, large reworks of the program’s scripts would need to be done in order to accommodate the new data sources. This was the key trigger that lead to the decision to build a new AMS Monitoring Interface.

The new AMI described in this paper is a state-of-the-art monitoring interface that combines the key features of the previous software into a unified system. The goal was to build a fast and modern system using the new tools and design philosophies developed in recent years. It efficiently manages the large amounts of data generated by the AMS, provides a noticeable speedup in plot generation compared to the legacy AMI, uses resources efficiently, allows for scalability and future-proofing, and enables an easier way to add new data sources without needing large reworks in the future. In addition to health status, it also supports access to scientific data such as trigger and detector performance, including calibration, occupation, and noise levels amongst other statistics, eliminating the need for most additional monitoring console programs. Finally, it is designed to have all features be fully accessible via a transportable user interface, allowing complete remote monitoring.

This paper discusses the design philosophies and software used to develop this new AMS Monitoring Interface.

\section{AMS Monitoring Data}

\subsection{Data Delivery}

\begin{figure}[htb]
    \centering
    {\includegraphics[width=0.7\linewidth]{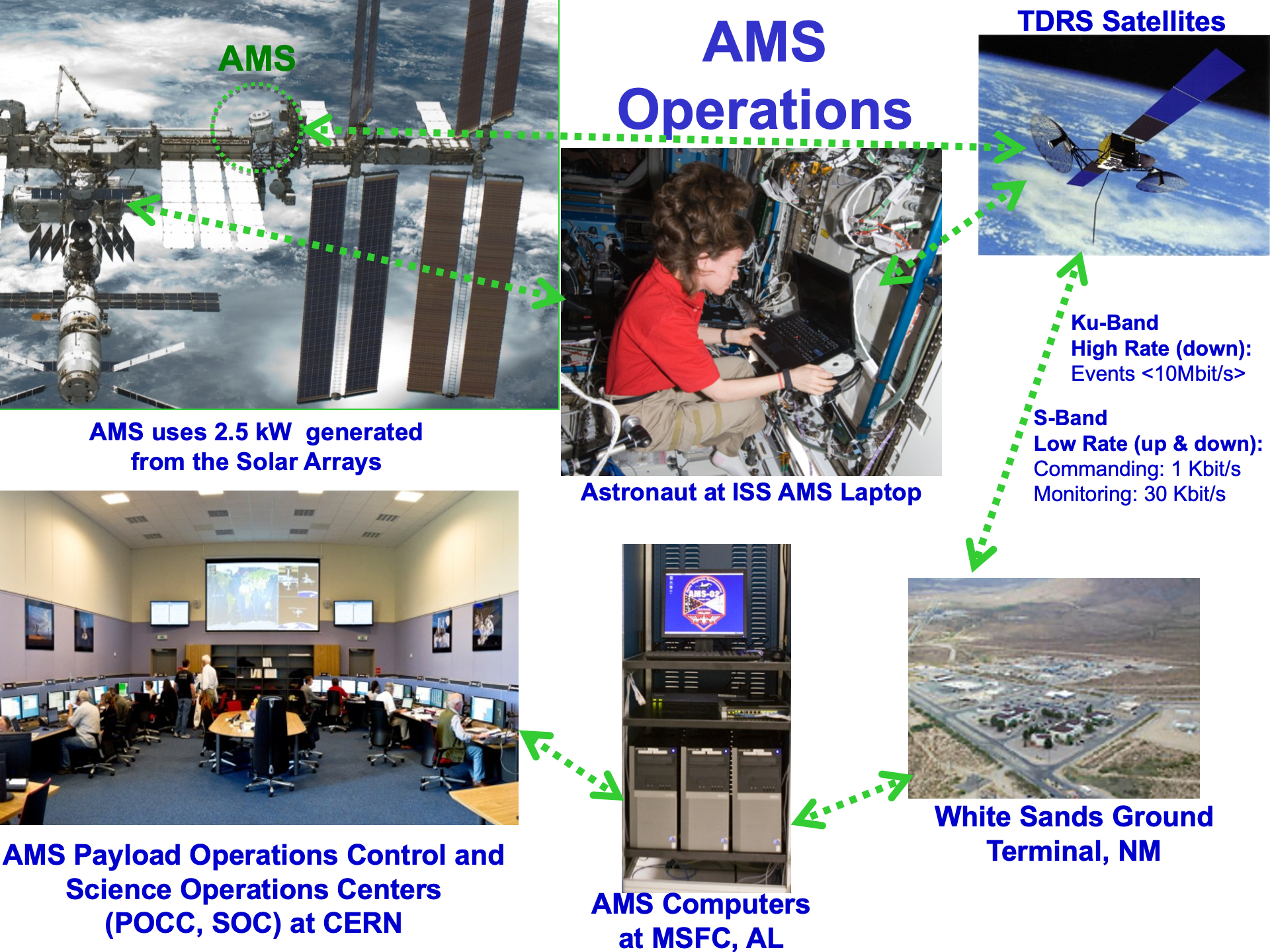}}
    \caption{Overview of AMS Operations. The data is sent down following the path clockwise from the upper left, while commands from the POCC follow this path in reverse (counterclockwise, from the lower left).}
    \label{fig:operations}
\end{figure}

Figure \ref{fig:operations} shows an overview of the flight and ground control operations carried out at the POCC. After being collected, monitoring data is streamed to the ground via the High Rate Data Link (HRDL) to ISS avionics. The ISS avionics devices then send this data to the Tracking and Data Relay Satellites (TDRS), using the Ku-band antenna, which then transmit it to ground at White Sands, NM, where it is relayed to AMS computers at NASA’s Marshall Space Flight Center (MSFC), AL. From there it is then finally delivered to the AMS Payload Operations and Control Center (POCC) at CERN, Geneva, Switzerland.

To guarantee reliable delivery from the ISS to the ground, as well as to achieve real-time delivery for some of the monitoring data, the data is split into two main streams, real-time and buffered. The real-time stream's bandwidth is low and mostly contains data from the various AMS sensors. The buffered stream has a larger bandwidth and contains both this sensor data and all of the science data. However, it is not real-time and has a time delay. A copy of all the data is also stored on the AMS laptop onboard the ISS, for backup purposes in case the ISS Ku-band antenna is not locked on a TRDS satellite or there is data lost during the high rate data link transfer.

At POCC, the monitoring data on the ground appears as tree-like hierarchy of nodes and unique data types associated with each node. AMS was built with a 4-fold redundancy, so there are many similar, but independent, sensors onboard. As such, they are categorized using sensor type (denoting measurement type such as temperature), sensor ID (distinguishing between sensors of the same type), and sensor name (unique names for each of the sensors). All streamed data is stored at the POCC servers at CERN as binary files, called AMS Raw Data, and ordered into file structures by time.

\subsection{Data Types}

Different data types describe the different measurement results from the detector. Alongside science data collected from the subdetectors, onboard sensors collect various health data measured from these subdetectors and their electronics. This health data, called housekeeping data, includes temperatures, pressures, pump speeds, voltages, currents, and data transfer rates. When taken together with select information from science data used to help with monitoring, such as calibration data, occupancy levels, and trigger rates, it is collectively referred to as monitoring data.

\begin{figure}[htb]
    \centering
    {\includegraphics[width=0.80\linewidth]{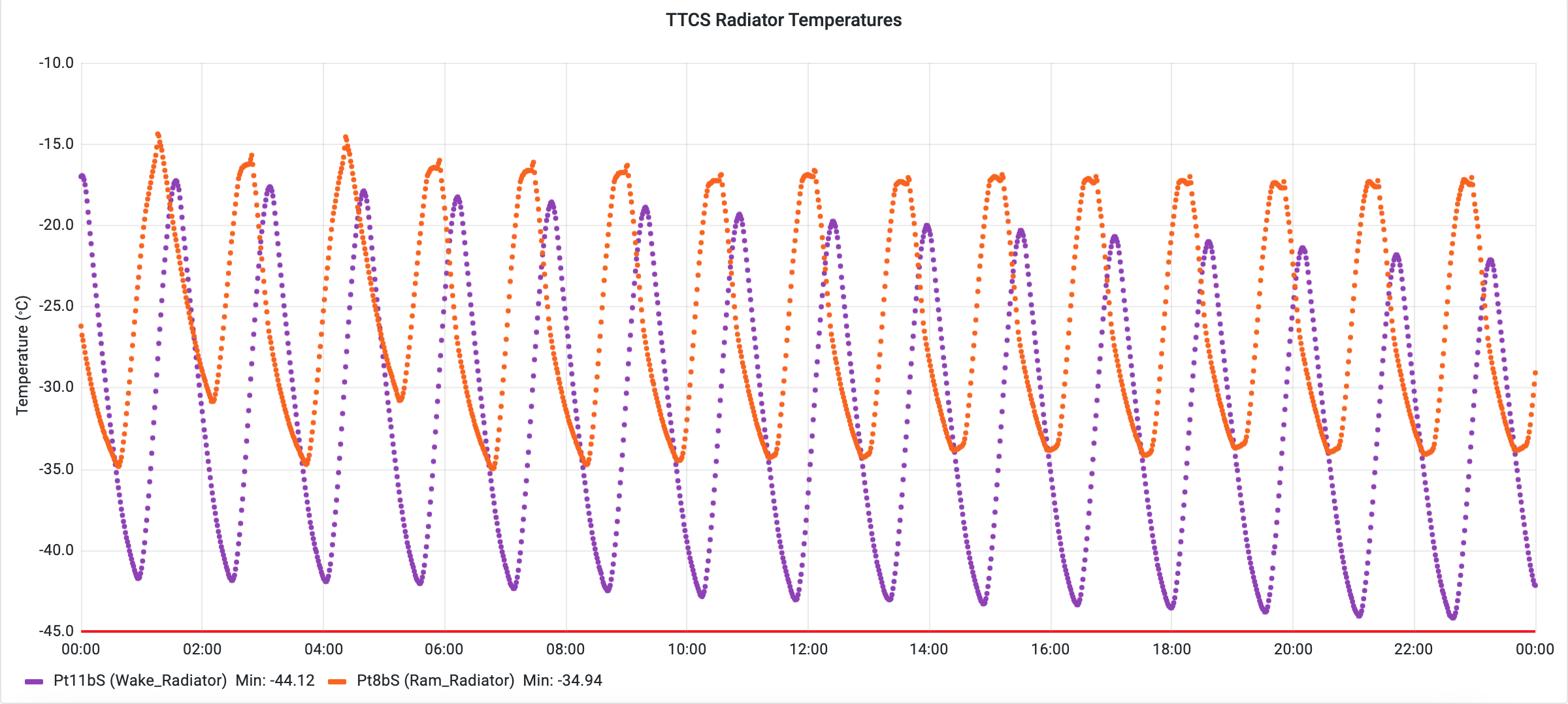}}
    \caption{Variation over the course of 24 hours on 6 December 2021 for two sensors measuring the temperatures at the center of the tracker radiators. The sensors on the Wake radiator (Pt11bS) and Ram Radiator (Pt8bS) show a variance of approximately 23.25 °C and 17.37 °C throughout a single ISS orbit, respectively.}
    \label{fig:temp}
\end{figure}

AMS temperatures are measured by 1118 temperature sensors plus an additional 78 sensors added during the UTTPS upgrade. Their purpose is to ensure that critical hardware temperatures stay within certain ranges of  acceptable values. Due to the AMS’s constant exposure to radiation, extreme temperature variations, changes in ISS orbit and orientation, solar effects, and docking/undocking of spacecraft, temperature fluctuations must be constantly monitored so that POCC can assess situations immediately and take action if necessary. Various subsystem cooling/heating mechanisms can be adjusted to regulate temperature. Figure \ref{fig:temp} gives an example of a temperature variation throughout 24 hours for the sensors measuring the temperatures on the outside of the tracker radiators, Pt8bS and Pt11bS, using one of the new AMI's Grafana panels.

In addition to temperature sensors, there are a number of voltage and current sensors that ensure large variations in values do not damage the electronics and power distribution systems. Other housekeeping data generated includes pressure sensors, such as for the TRD subdetector (CO2 diffusion can affect signal generated for events \cite{Heil2013a}) where the pressure within the tubes is constantly monitored to determine when CO2 should be resupplied. Another instance is within the UTTPS, where CO2 liquid to vapor percentage is monitored to ensure optimal cooling. The UTTPS has additional monitoring as well, such as the pump rotation speeds to ensure optimal performance. If the running pump suddenly slows, experts at POCC should be alerted to take action and find the cause of the slow down, preventing possible permanent damage.

In addition to the housekeeping data types mentioned, subdetector specific housekeeping data, firmware state, and data rates are also monitored and sent to ground with a time granularity on the order of one second to few minutes. This data then needs to be processed, visualized and analyzed by the AMS Monitoring Interface.


\subsection{Addition of UTTPS}

Electronics for the central part of the AMS Silicon Tracker (six out of nine layers) produce approximately 144 W of heat which need to be withdrawn and dissipated into space \cite{VanEs2009}. The original Thermal Tracker Cooling System (TTCS) was a two-phase CO2 system and was mechanically pumping CO2 in its liquid phase. It was designed to maintain the Tracker temperature between -20° C to +20° C. There were four redundant pumps and two redundant sets of pumping loops and control electronics.

In 2014, the TTCS reached its expected lifetime \cite{Mussolin2020OverviewCampaign}. In order to continue functionality and improve the cooling of the Tracker, a new Upgraded Thermal Pump System (UTTPS) was designed and built. Towards the end of 2019 and early 2020, NASA arranged for four extravehicular activities and the UTTPS was successfully installed. 

The UTTPS added an additional 78 temperature sensors and 6 absolute pressure sensors. Alongside this, there are 3 hull sensors on each pump to measure the RPM and feedback from the pump controller electronics is monitored for the voltage and current measurements. Adding these measurements required a complete rework of the existing AMS Monitoring Interface.

\section{Legacy AMS Monitoring Interface}

The legacy AMI \cite{Alberti2011}, which was operated until the end of 2019, was a collection of monitoring consoles and a Relational Database Management System. The relational database was used as the backend, a Web2py Database Abstraction Layer (DAL) was used to hold the database structure description, a Remote Procedure Call (RPC) interface allowed client programs to interact with the database, an AMI web interface showcased RRDtool \cite{TobiasOetiker2017AboutRRDtool} generated static plots, and a Scanner utility filled in the database using the RPC interface.

\begin{figure}[htb]
    \centering
    {\includegraphics[width=0.7\linewidth]{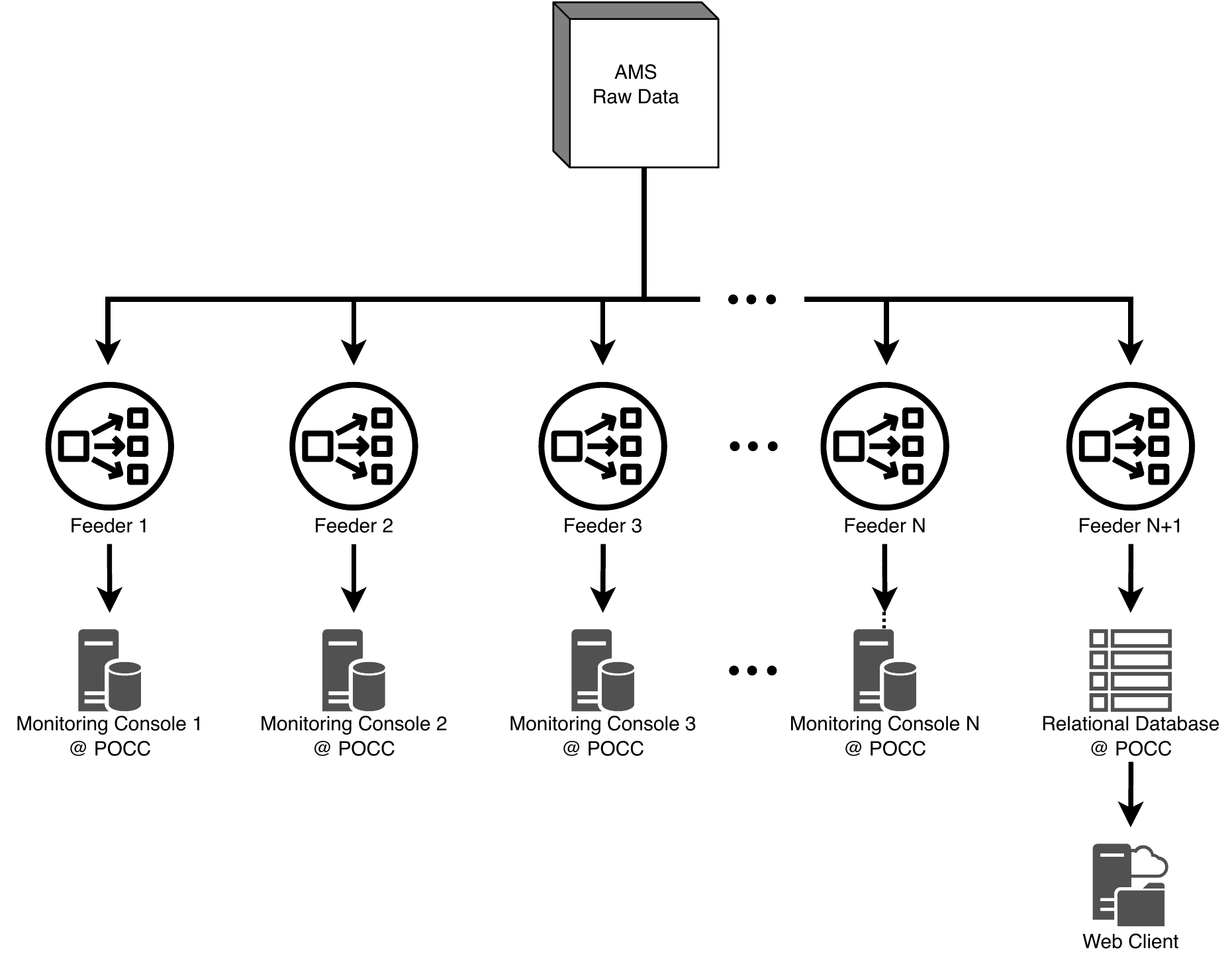}}
    \caption{Structure of the legacy AMS Monitoring Interface. Each monitoring console on the POCC had its own dedicated feeder which would parse through the Raw Data to get the information the console needed. One of the feeders was dedicated to feeding the relational database, which powered the old AMI's web client.}
    \label{fig:old_ami_outline}
\end{figure}

\begin{figure}[htb]
    \centering
    {\includegraphics[width=0.7\linewidth]{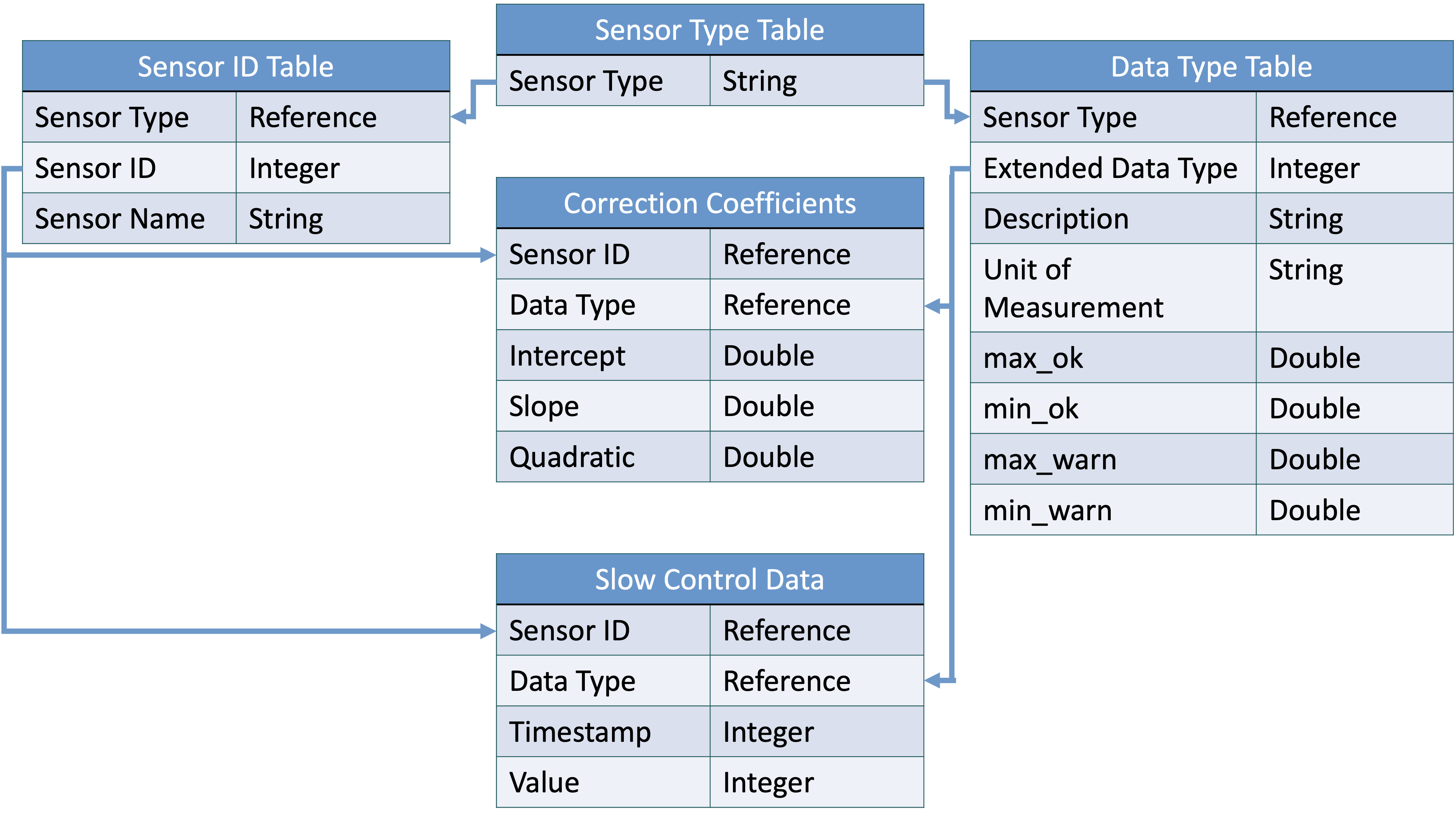}}
    \caption{Simplified layout of the legacy AMI’s Relational Database Management System, adapted from \cite{Alberti2011}.}
    \label{fig:old_ami}
\end{figure}

Figure \ref{fig:old_ami_outline} shows an outline of the legacy AMI's structure. The feeder utility, made using C++, would run individually for each of the monitoring consoles at POCC, with an additional one being run for the relational database. It scanned AMS Raw Data and optionally a single stream of data via the POCC multicast. It searched for address numbers and determined the sensor type, sensor name, sensor ID, and data type before feeding it into the relational database using the RPC interface. A simplified layout of the relational database is shown in Figure \ref{fig:old_ami}.

The web interface allowed the user to browse, by name, the extended  data type table for a given sensor type and select a time window to query the data points within that time frame. A custom Python script within Web2Py would then extract the requested data from the relational database and use RRDtool to generate a plot of it, with respect to the time, as a PNG image. This was done completely on the server side with the images being uploaded to the web interface. Certain pages on the web interface could show multiple plots either within a single image or with different images in order to show relationships and give a comprehensive view of a subsystem. Such multi-plot pages were made by modifying the Python script and the DAL interface was used to develop an internal set of functions that would grant access to the data. Additional monitoring consoles were only accessibly physically inside the POCC and not on the web interface.
 
Due to the heavy reliance on scripting, any kind of edit or new plot would require the maintainer of the interface to modify the code. In addition, the reliance upon the Relational Database Management system and the constant parsing through the AMS Raw Data showcased an outdated model of monitoring critical AMS health data. Parsing through the Raw Data files multiple times for each console, storing it on various tables in a relational database, creating static PNG images upon refresh took a considerable toll on computing resources, was slow, and overall inefficient. Much of the processing, both from the backend (retrieving the data) and frontend (generating the plots to show the data) took place on the server side, which further represented an outdated model for monitoring data. 

Additionally, implementing certain functions such as extracting data points as a CSV file for further processing, emailing warning alerts when temperatures were reaching their limits, or building heatmaps for certain housekeeping data needed custom monitoring consoles to be built each time. Furthermore, the UTTPS upgrade brought new measurements, which required massive rewrites in the scripts to integrate the data into pre-existing pages and create entirely new pages on the web interface.

The UTTPS upgrade, along with the general slowness and inefficiency of the legacy AMI model, prompted the need for an upgrade to the AMS Monitoring Interface system.

\section{New AMI System Structure}

\begin{figure}[h!]
    \centering
    {\includegraphics[width=0.35\linewidth]{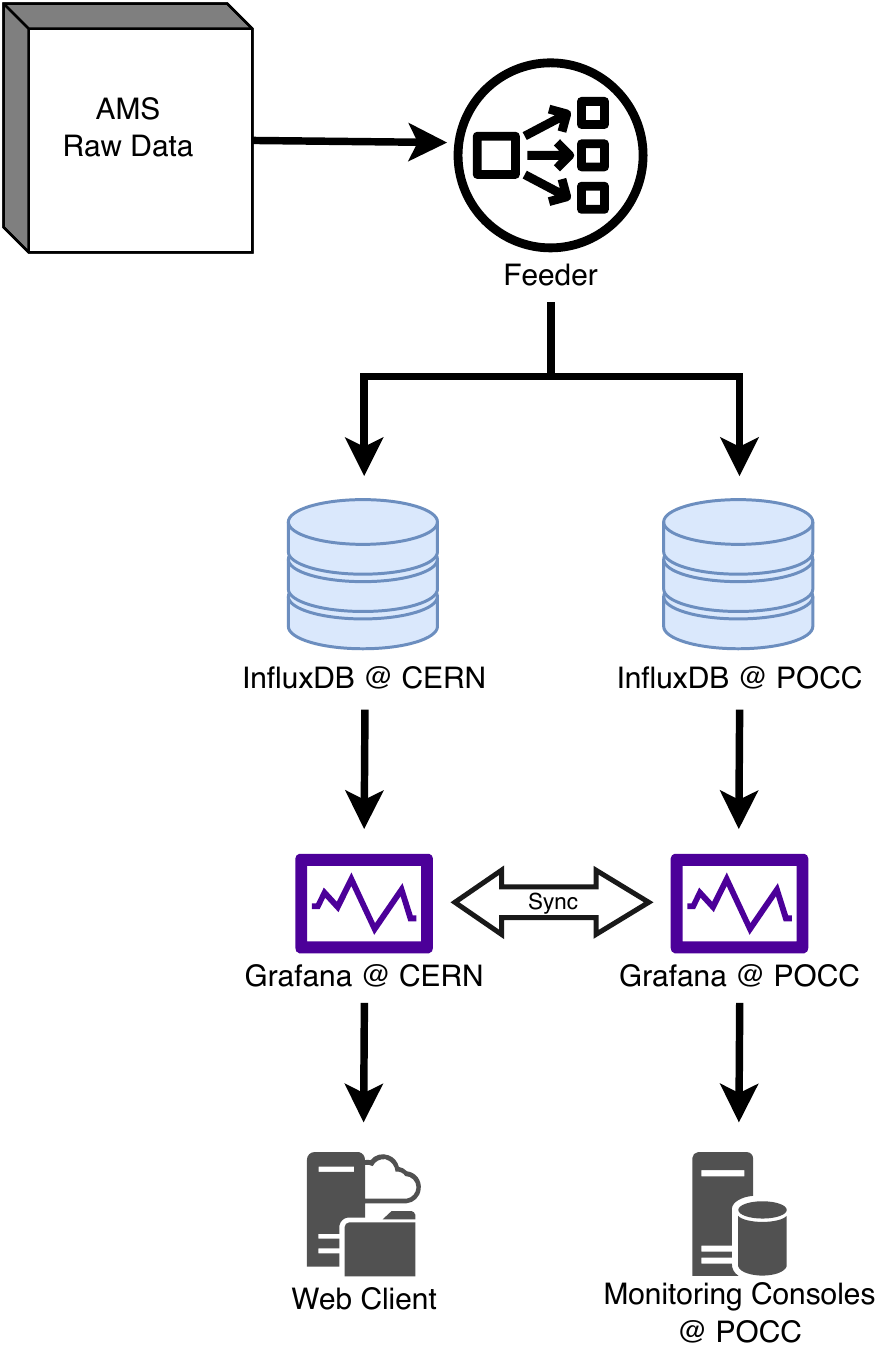}}
    \caption{Structure of the new AMS Monitoring Interface. The Raw Data is passed to the feeder, which then fills two InfluxDB databases, one stored on the CERN servers, one stored on the POCC's servers, for redundancy. Both have independent Grafana instances that display the data, with both Grafanas using Git to synchronize once per day. The CERN instance can then be accessed by the internet for the general AMS members while the POCC instance is used for the monitoring consoles at POCC. }
    \label{fig:new_ami_outline}
\end{figure}

Figure \ref{fig:new_ami_outline} outlines the structure of the new AMI. The new AMI uses the open-source, publicly available, time series database InfluxDB as its backend and the time series data analytics software, Grafana, as the frontend. It is designed such that Raw Data is parsed only once by the feeder and then subsequently passed to two independent InfluxDB databases, one hosted on CERN's Database on Demand (DBOD) \cite{Aparicio2012DataBaseDemand} servers and an identical one hosted on POCC's servers, for dual redundancy. Two Grafana instances are then independently connected to each of the InfluxDBs, but the plots and dashboards are kept the same via Git synchronization occurring once a day. The CERN version of Grafana is hosted on the OpenShift Platform \cite{Lossent2017PaaSOrigin} while the POCC version is hosted on the POCC servers. The CERN version therefore acts as a web client for the general AMS members while the identical POCC version is used as a replacement for the monitoring consoles at POCC.

\section{Backend Design}

\begin{figure}[htb]
    \centering
    {\includegraphics[width=0.5\linewidth]{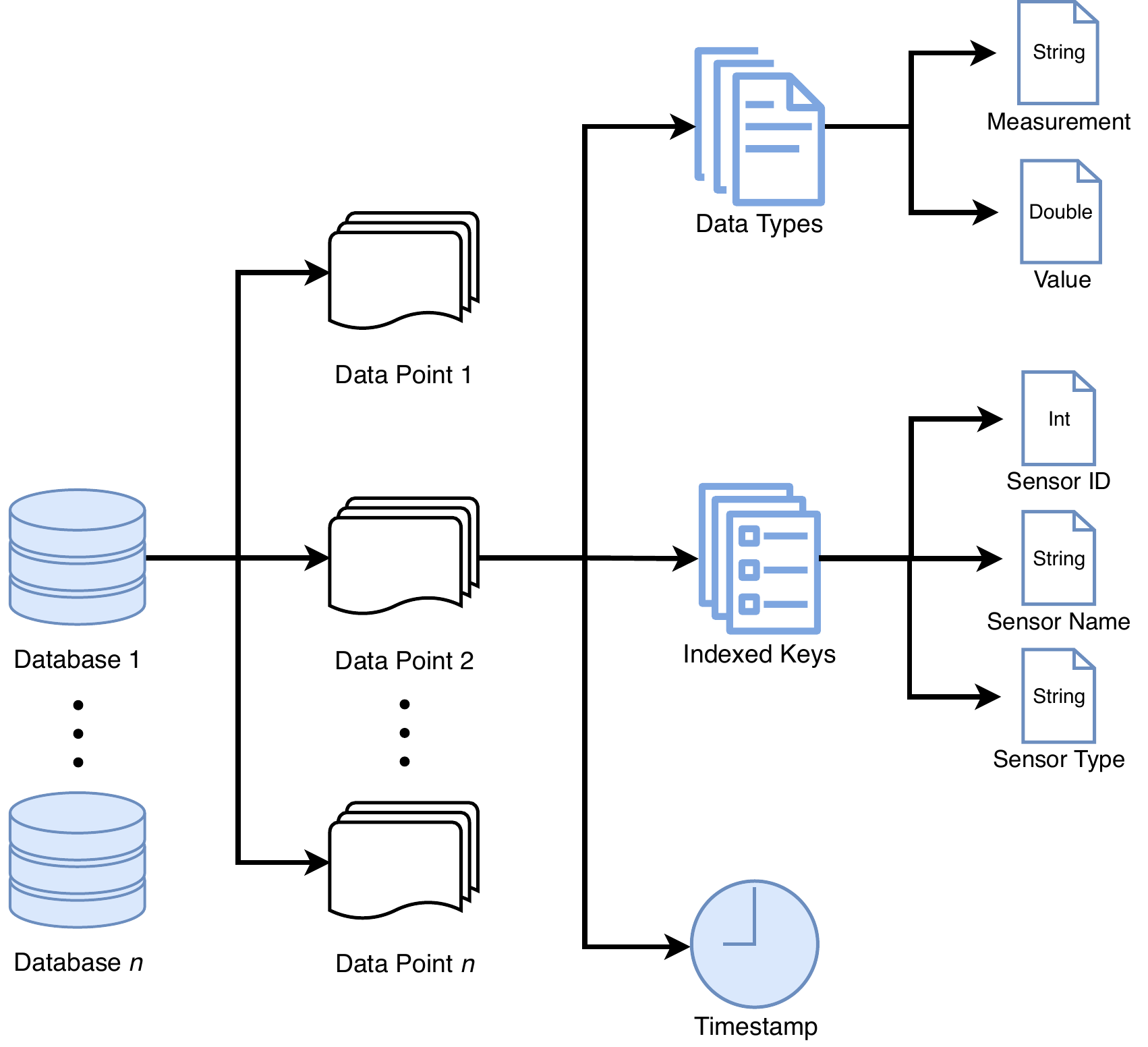}}
    \caption{Simplified outline of the new AMI InfluxDB database. A single database or multiple databases can be used simultaneously. Each of these databases contain time series data in the form of data points, which have 3 major categories: Data Type which includes the measurement name (for example, temperature) and value, the corresponding timestamp, and indexed keys, which are flexible tags that allow the querying and displaying of entire groups of data points, such as "all temperature sensors".}
    \label{fig:New_AMI_Overview}
\end{figure}

A new feeder program was developed using C that can parse through AMS raw files or a data stream via the POCC's multicast, identify key sensor IDs, extract sensor data, sensor type, data type, their respective values and timestamp, and send the values to the InfluxDB database. Figure \ref{fig:New_AMI_Overview} shows an outline of the new AMI database structure.

The C program routinely scans for new raw files, with each file representing one minute of the data streams, and parses through them to find the relevant data. It sets the field key to be the measurement, such as temperature, voltage, or pressure, with the field value being the corresponding measurement’s value, and sets various key tags identifying the name, id, or type of measurement.

To send a data point with accompanying timestamp, field, and keys, InfluxDB’s HTTPS line protocol is used with the CERN and POCC databases listening for https requests on select ports. The C program creates a string with the relevant information, identifies which database server to send the data to, takes in credentials and a port number and finally uses the line protocol to submit this information. The database then receives it and stores the information.

Once this process is complete, the raw files no longer need to be processed and are kept in storage. Data from the database can then be accessed directly via commands to the database by users at POCC, or via a time series data analytics software (see Section \ref{section:Frontend}).

\subsection{Benefits of using a Time Series Database}
A time series database is optimized for storing data points indexed with an inherent temporal order, otherwise known as a time series, with associated pairs of times and a corresponding value generated at that time, as formalized in 2002 \cite{lonardi2002finding}. Since then, time series and related databases have been used in a variety of fields including medicine \cite{abe2005implementing}, weather prediction \cite{mcgovern2007understanding, McGovern2010IdentifyingPrediction}, computer generated graphics \cite{celly2004animated}, and even computing resource management at ATLAS \cite{Beermann2020ImplementationGrafana}.

With relational databases, data with similar features are organized into tables and linked with relationships, as shown with the legacy AMI. They are most efficient when data points have large correlation with one another and there is a need to edit data points after insertion. AMS monitoring data, however, has little need to be edited after insertion and has high cardinality. This causes relational databases to perform poorly, since long tables can no longer be cached resulting in data querying taking significantly long periods of time.

With the time series data format, however, the issue of cardinality is reduced because the format is optimized to handle data points with time stamps. Time series data have natural temporal ordering with relationships defined by simple time measurements, data values, fields describing data types, and keys adding metadata. While this limits the type of relationship data points can have, it simplifies the database structure layout (see Figure \ref{fig:New_AMI_Overview}) and increases efficiency by allowing chunks of data, or shards, within a selected time frame to be located and prepared for querying, allowing faster information retrieval \cite{Fadhel2018AData}.

Additional benefits of time series data include the ability to aggregate and downsample data points over time, track and monitor time series, process series for further tasks, perform various math operations (depending upon the exact time series database used), export data for further processing (see Section \ref{section:simple_data}), and display them using various monitoring tools. 

Due to the ease in which the database can be filled, the addition of future new measurements of data, similar to how the UTTPS upgrade brought additional measurements, is a straightforward process. Large rewrites of the scripts are no longer necessary, and instead, simple additional clauses in the feeding scripts can allow it to search for additional new measurements and add them to the database with corresponding field-key pairs and metadata tags. 

Treating housekeeping and certain science data as a time series puts the emphasis on the timestamp of a datum. This allows the seamless merging of the different data streams from the ISS, which originally start as a single stream. The smaller and faster real-time stream is processed first to fill in the database while any gaps between data points are filled later as the slower, buffered stream arrives and is processed. Additionally, duplicated data points can easily be overwritten based on their timestamp, negating the need for an extensive search of the original data. Finally, time series database structure allows users to query only points of interest, selected by the field-key pairs and metadata tags and bounded by the time range chosen.

This method reflects a more up-to-date form of processing and storing monitoring data.

\subsection{Resources Benefit}
With the new feeding mechanism and time series database, AMS raw files are processed only once to extract the relevant data points to store within the dual redundant databases. Additional AMS raw files do not require the processing of previous files to get more cohesive outputs, unlike the legacy AMI. This significantly reduces the number of times the large raw files are parsed, reducing load on the servers and preventing the need for storing the large raw files on the server, saving hard disk space.

\subsection{Benefits of InfluxDB over Alternatives}
InfluxDB is an open source time series database written in the Go programming language with no external dependencies \cite{InfluxDBInc.2021InfluxDBDocumentation}, allowing for the efficient development of custom pre-compiled binaries and easy deployment on various servers and operating systems. It uses a custom-built Time-Structured Merge Tree storage engine which compresses series data in a columnar format, storing only the differences between values in a series. This makes it highly efficient for both storage and querying of data and allows AMS to effortlessly store half a terabyte of raw data per year on a single server. It uses an SQL-like query language, making it simple to retrieve data, and supports a large number of aggregation and math functions during querying \cite{InfluxDBInc.2021InfluxDBDocumentation}.

When compared to other time series databases, previous studies have shown InfluxDB to be far simpler to use and more efficient even on low powered devices \cite{Fadhel2018AData}. While InfluxDB does not have the fastest insertion time, it has the fastest data querying compared to competing time series databases such as TimescaleDB \cite{Grzesik2020ComparativeNetworks}. This is more advantageous for the AMS Monitoring Interface as inserting data has been reduced to a single pass per AMS raw file, but data is constantly queried every few seconds. Finally, InfluxDB has consistently ranked as the most popular time series database on DB Engines rankings, outshining competitors like Prometheus and Graphite \cite{DB-Engines2021DB-EnginesDBMS}.

\section{Frontend Design}
\label{section:Frontend}

The new AMI was designed with the intent to build one frontend to eventually replace the various monitoring consoles at POCC. Grafana was chosen as a suitable candidate due to its focus on time-series data analytics, support for various plugins, and ability to display data in a variety of formats (see section \ref{section:Implementation}).

\subsection{Benefits of using Time Series Data Analytics Software}
Plots and similar visualizations are used to better understand the monitoring data, make comparisons, understand trends, and spot abnormalities. Grafana is the open source, web-based, interactive visualization and dashboard data analytics software selected for this task \cite{GrafanaLabs2021GrafanaFeatures}. It has built-in support for InfluxDB, along with many other database software, and allows for easy reading of their data without the need for programming.

While the legacy AMI relied upon a single script manager to modify the plots and pages that organize them, due to the simple web interface nature, Grafana can be modified by the detector experts of AMS, without requiring programming experience. This substantially improves update speed since changes to a particular subdetector’s plots can be tasked directly to the team that manages the monitoring of said subdetector. This improves usability as teams can create dashboards to fit their individual needs and allows for an overall collaborative, concurrent design of the AMI. This prevents responsibility from falling on a single person, eliminating the risk of a single point of failure. 

Furthermore, unlike the legacy AMI, the new Grafana-based AMI allows for the export of data points into a CSV text format using the web interface itself, directly from the panels. While such exporting is possible directly from the InfluxDB backend, it often requires the use of its own InfluxDB command line interface. Furthermore, it requires direct access to the backend, which is limited to only the database administrators, for security and stability reasons. 

The ability to export data points allows a wider access to the data than ever before, with members no longer needing proprietary AMS software or bulky AMS raw data to work with monitoring data.

\subsection{Resource Benefits}
While the legacy AMI followed an outdated strategy of developing the visualizations on the server side and only displaying them for the user, Grafana follows the more modern approach of letting clients handle the visualizations.

Grafana efficiently uses client resources to query data and create plots, which significantly reduces the load on the server as each page refresh no longer causes the server to parse a relational database to build new plots. Instead, the server simply streams the time series data points while Grafana on the client side takes care of the plotting and visuals. Since only data points are streamed, the bandwidth is small and allows for fast access and refreshes on the new AMI.

Additionally, due to Git-powered syncing, both Grafana instances are updated when one is modified, reducing maintenance cost in terms of manpower.

\subsection{Benefits of Grafana over Alternatives}

Grafana allows the building of complex plots and supports a variety of display devices, including mobile devices, without requiring any additional programming. On the client side, since Grafana uses a web interface, clients do not install anything and can simply point their web browser to the Grafana server URL to access the AMI. This allows for easy access by all devices and operating systems, without the need for any setup.

Grafana’s simplicity, easy accessibility from multiple devices, and visually attractive plots made it the ideal choice for the AMI’s time series data analytics software. Its ability to allow collaborative design by authorized users via the web interface enables different monitoring teams to customize their dashboard immediately themselves without the need for the original programmers to modify the script.

\begin{figure}[htb]
    \centering
    {\includegraphics[width=0.80\linewidth]{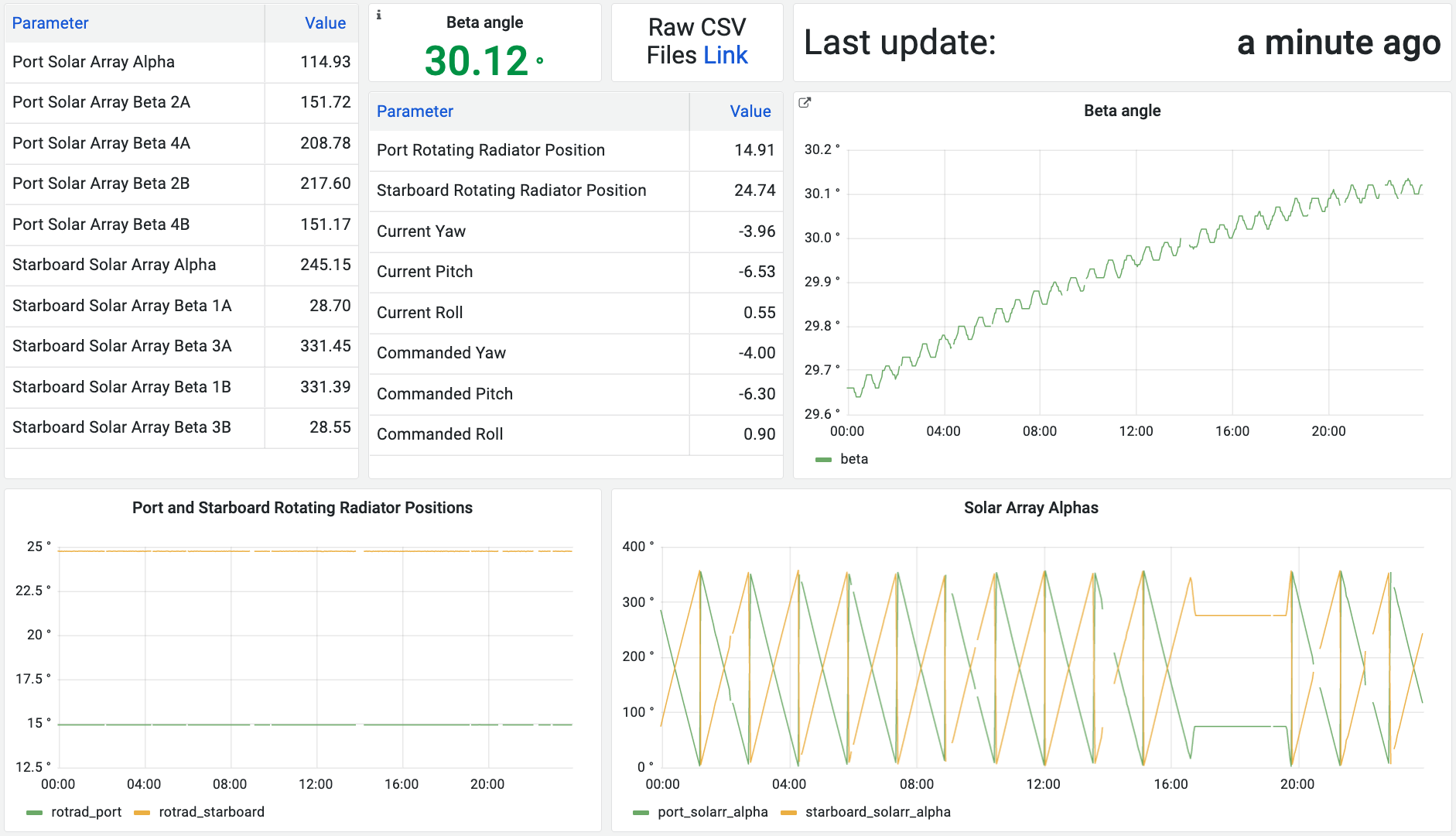}}
    \caption{An example of a dashboard commonly used at the POCC. It shows the usage of single value stats (the current Beta Angle and the Last Update panel), tables (showing ISS rotation and solar array parameters), scatter plots (such as beta angle), and a custom-made panel (direct downloading of a CSV for the data).}
    \label{fig:iss_page}
\end{figure}

Grafana supports a wide range of visualization tools such as various charts, heatmaps, histograms, pie charts, gauges, tables, single value stats, node graphs, and more. An example of some of these types of plots can be seen in Figure \ref{fig:iss_page}, which showcases a typical dashboard used in the new AMI (additional examples can be seen in Section \ref{section:Implementation}). In addition, Grafana has custom plug-in support so missing features can be programmed in. It supports a wide range of client libraries making its API accessible via Python, JavaScript, Go, Arduino, and more, making it versatile. Finally, it has support for custom alerts and alarms that can warn predefined users if a value goes above or below a certain threshold via email, SMS, or instant messengers \cite{GrafanaLabs2021GrafanaFeatures}.

The above outlined features allow shift takers to coalesce their different monitoring software into a single Grafana interface due to the wide range of supported visualizations, alerting methods, and support for custom plug-ins. This streamlines the job of shift takers and improves efficiency and response time. 

Additional benefits of Grafana over competitors, such as Graphite’s own visualization tool and Kibana, are the native support for InfluxDB and additional support for simultaneous querying of data from different databases. Grafana’s plug and play nature allows multiple database sources to be “plugged” in for easy querying. Replacing one database with another is simple and accessible via the web interface.

Lastly, the user interface is fully featured and customizable. It allows for simple analytics to be done instantly on the queried data without the need for the database itself to perform the analytics, features customizability for default time ranges, refresh speed, and viewing modes, and automatically scales to fit the screen it is being viewed on.

\subsection{Security Benefits}
While InfluxDB is restricted to database administrators, reducing security risks, Grafana is designed to be open to the AMS collaboration members across the globe. As such, Grafana’s innate security features are used to provide secure access to members.

Firstly, IP addresses and hostnames can be limited, blacklisted, or whitelisted to prevent specific sources from connecting if need be. Secondly, Open Authorization (OAuth) and Lightweight Directory Access Protocol (LDAP) was set up to allow CERN’s Single Sign On login for the former and enable AMS’s own LDAP authentication for the latter, each belonging to one of the two Grafana instances. Lastly, individual user permission can be limited to prevent unnecessary or accidental tampering. Currently, most users are “viewers”, shift designers who can modify panels and dashboards are “editors”, and admins of the AMI are appropriately titled “admins” with the ability to modify everything including security setups, backend databases, and more.

The increased security options and the unification of the previous monitoring systems into a single Grafana system had the additional bonus of allowing shift takers to monitor AMS health remotely. This was previously not possible with the legacy AMI as many monitoring software instances were only running on computers at the POCC. The universal web interface along with the up-to-standard security protocols allowed shift takers to perform their duties while away from the POCC, a feature that saw much use during the COVID-19 pandemic.

\section{Implementation}
\label{section:Implementation}

\begin{figure}[htb]
    \centering
    {\includegraphics[width=0.85\linewidth]{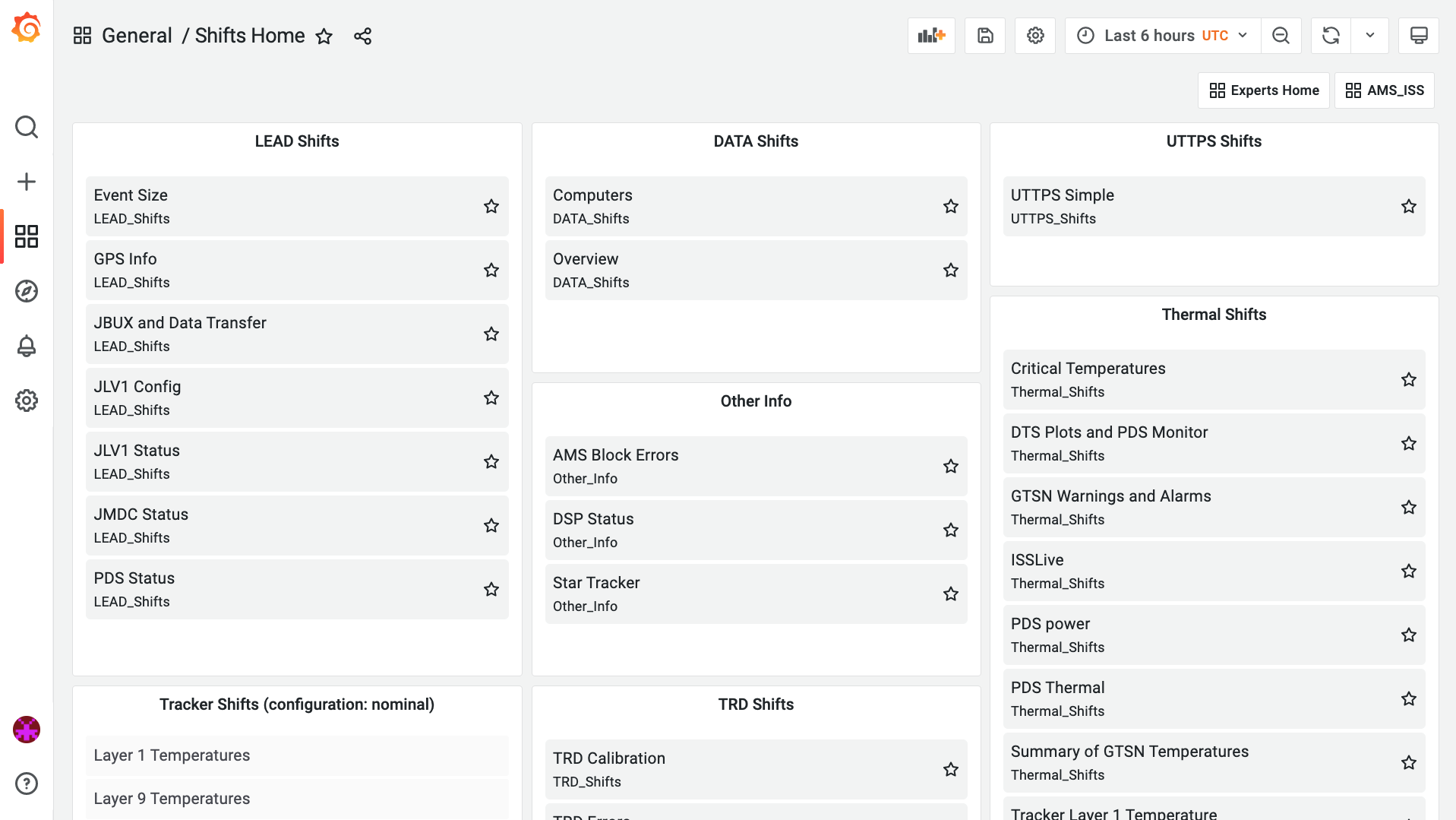}}
    \caption{Homepage of the new AMI, showcasing the different dashboards that shift takers at the POCC can quickly access.}
    \label{fig:homepage}
\end{figure}

\begin{figure}[htb]
    \centering
    {\includegraphics[width=0.7\linewidth]{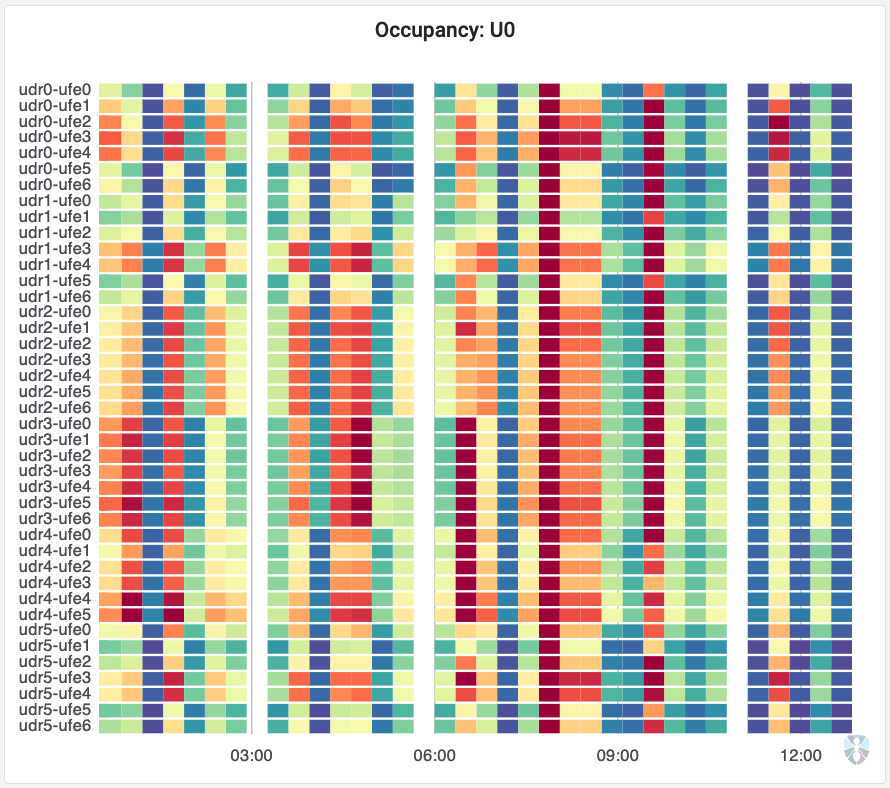}}
    \caption{Example of a status map being used to display the occupancy of the TRD. The occupancy of each channel can be seen for any given time. If a particular channel shows black, it indicates the channel has died. Conversely, if a channel shows the same color for a long period of time, it can indicate a large amount of noise is being consistently detected. Here, blue represents the lowest, nonzero, occupancy, while maroon shows the highest occupancy. Note: the status map was added via the statusmap plugin \cite{Flant2022StatusmapGrafana}.}
    \label{fig:trd_occupancy}
\end{figure}

\begin{figure}[htb]
    \centering
    {\includegraphics[width=0.7\linewidth]{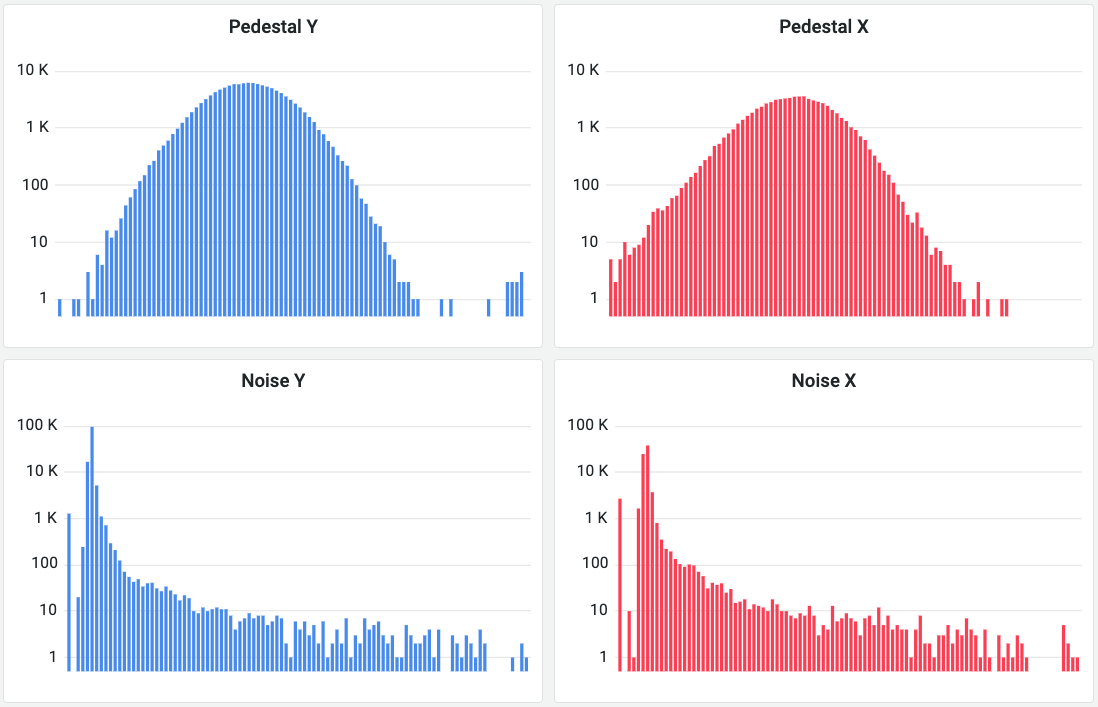}}
    \caption{An example of histograms. The Silicon Tracker is calibrated twice per orbit, when ISS crosses the equator. During this time, the pedestal and noise levels are measured for each channel, represented as histograms in this figure.}
    \label{fig:tracker_calibrations}
\end{figure}

\begin{figure}[htb]
    \centering
    {\includegraphics[width=0.7\linewidth]{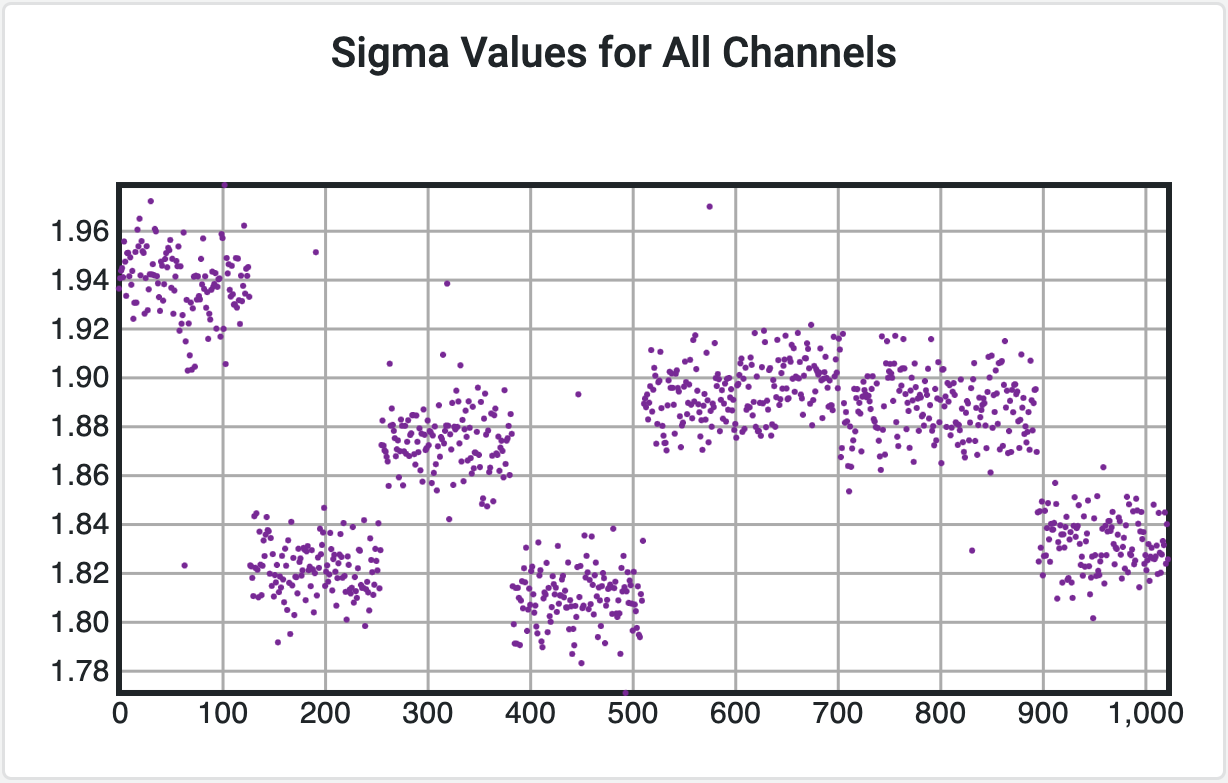}}
    \caption{An example of a panel that uses a silicon tracker layer's channels instead of time as the x-axis. This panel is a scatter plot used to showcase the sigma value for each of the layer's channel using the scatter plugin \cite{MichaelMoore2022Scatter}.}
    \label{fig:L0_Channels}
\end{figure}

\begin{figure}[htb]
    \centering
    {\includegraphics[width=0.7\linewidth]{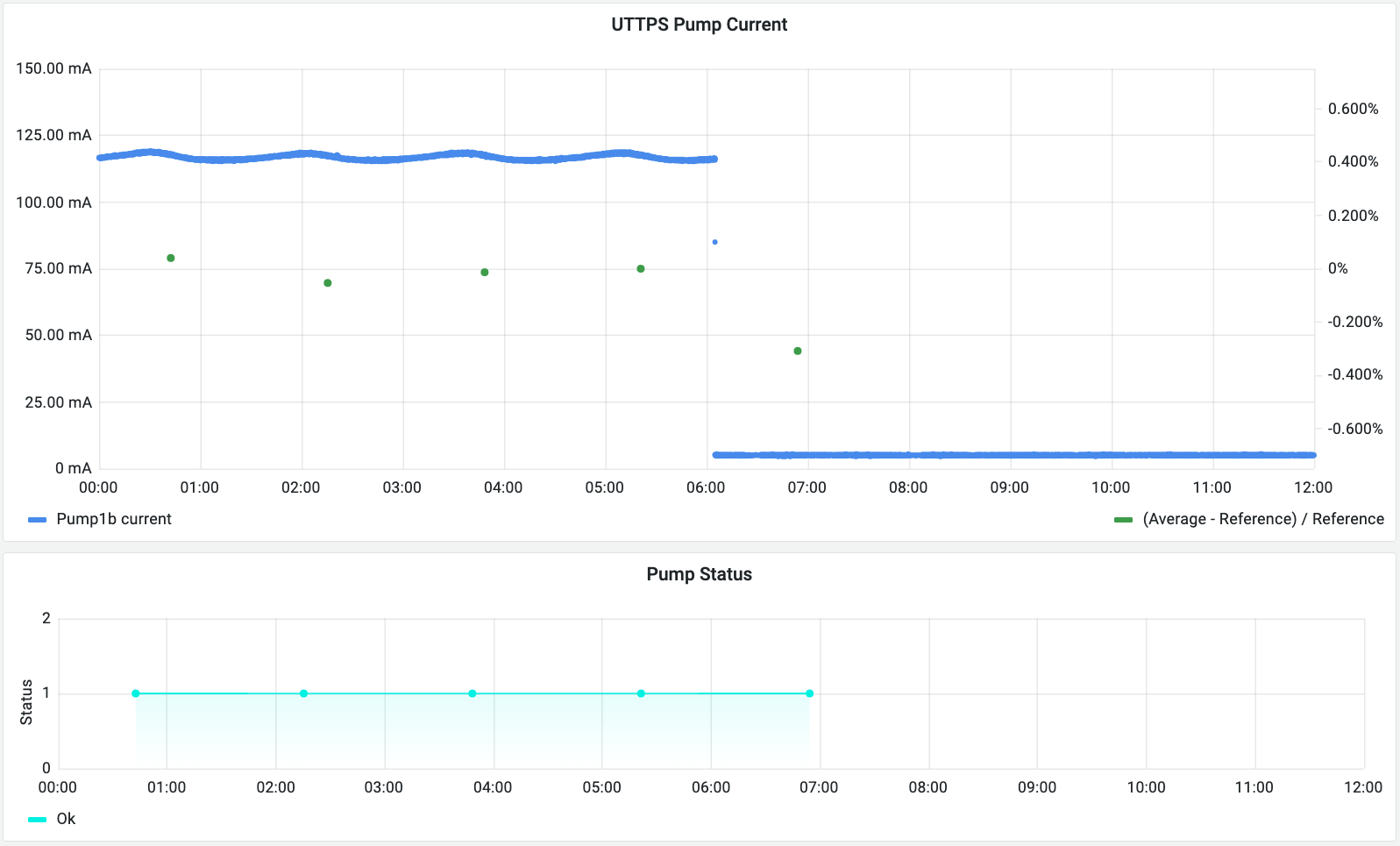}}
    \caption{An example of detector status monitoring, in this case the UTTPS, and basic math calculations. When the UTTPS is turned on, the pump current is displayed while the pump status is set to 1. We can see that, when the pump is shut off, the status and current become 0. The green points represent the weighted difference between the average and the reference current, an example of how Grafana can perform math on the data points directly, without it being a part of the database.}
    \label{fig:pump_status}
\end{figure}

\begin{figure}[htb]
    \centering
    {\includegraphics[width=0.8\linewidth]{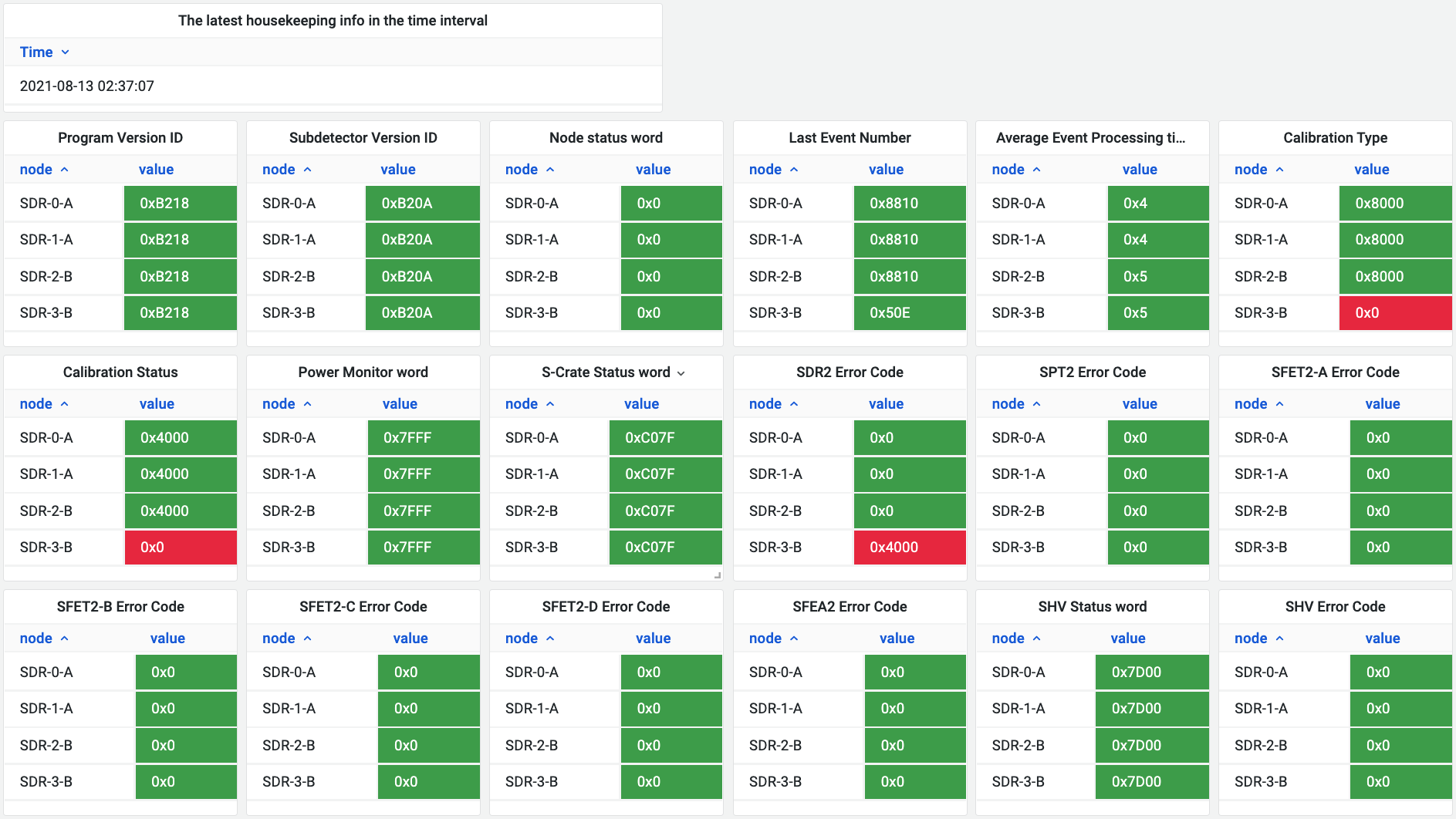}}
    \caption{An example of the table format being used as a status indicator. Whenever the ToF detector experiences a bit flip or reboots, certain entries show as red. When everything is nominal again, the status returns to green, allowing an at-a-glance look at the status of the detector.}
    \label{fig:tof_housekeeping}
\end{figure}

\begin{figure}[htb]
    \centering
    {\includegraphics[width=0.7\linewidth]{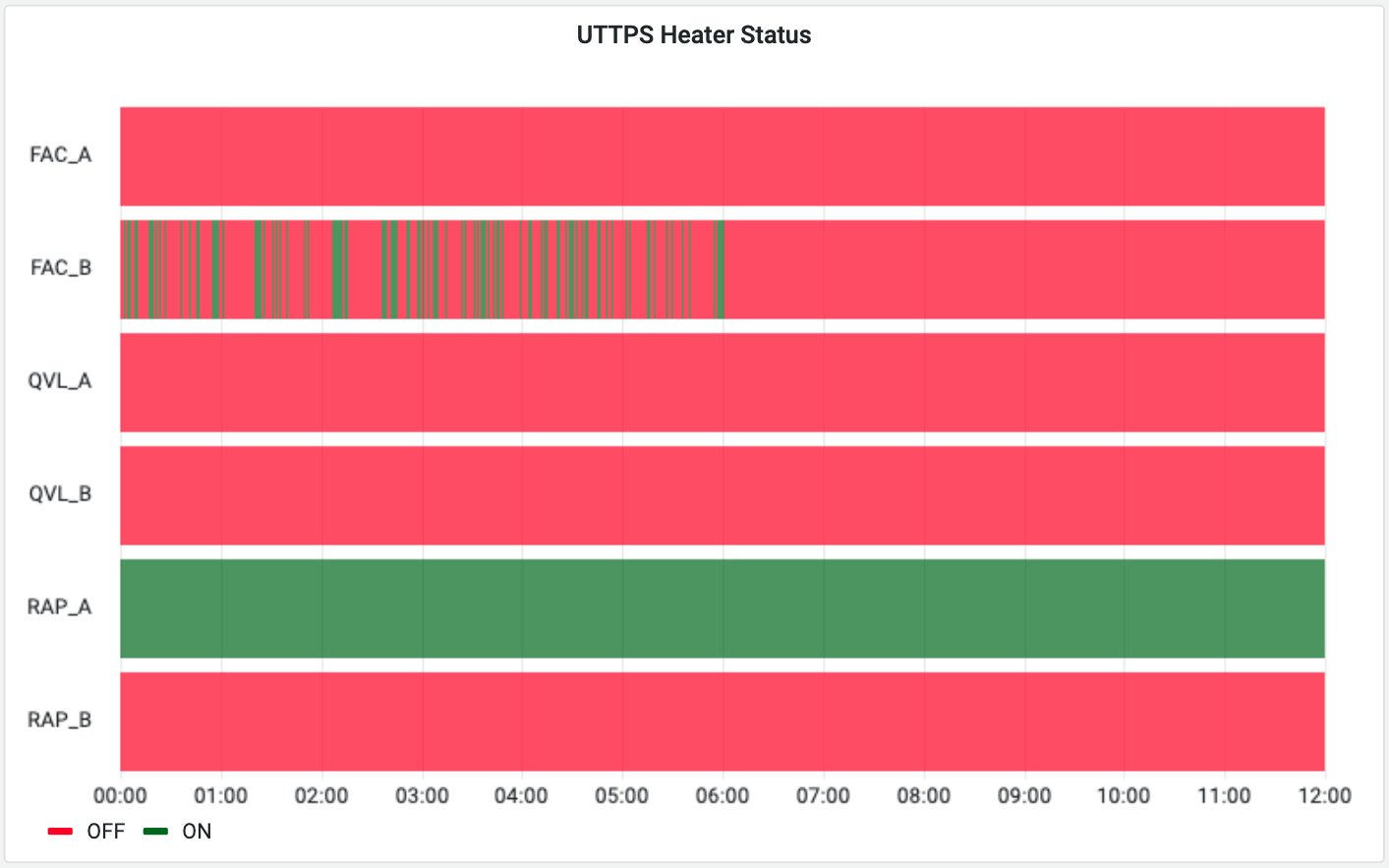}}
    \caption{An example of an on/off time indicator. The UTTPS has several heaters, six of which are shown here with their specific IDs on the left. At a glance, we can see that for the duration of 12 hours, RAP\_A was consistently on and FAC\_B alternated between on and off for the first 6 hours and then, like the other heaters, was off.}
    \label{fig:uttps_heater}
\end{figure}

\begin{figure}[htb]
    \centering
    {\includegraphics[width=0.8\linewidth]{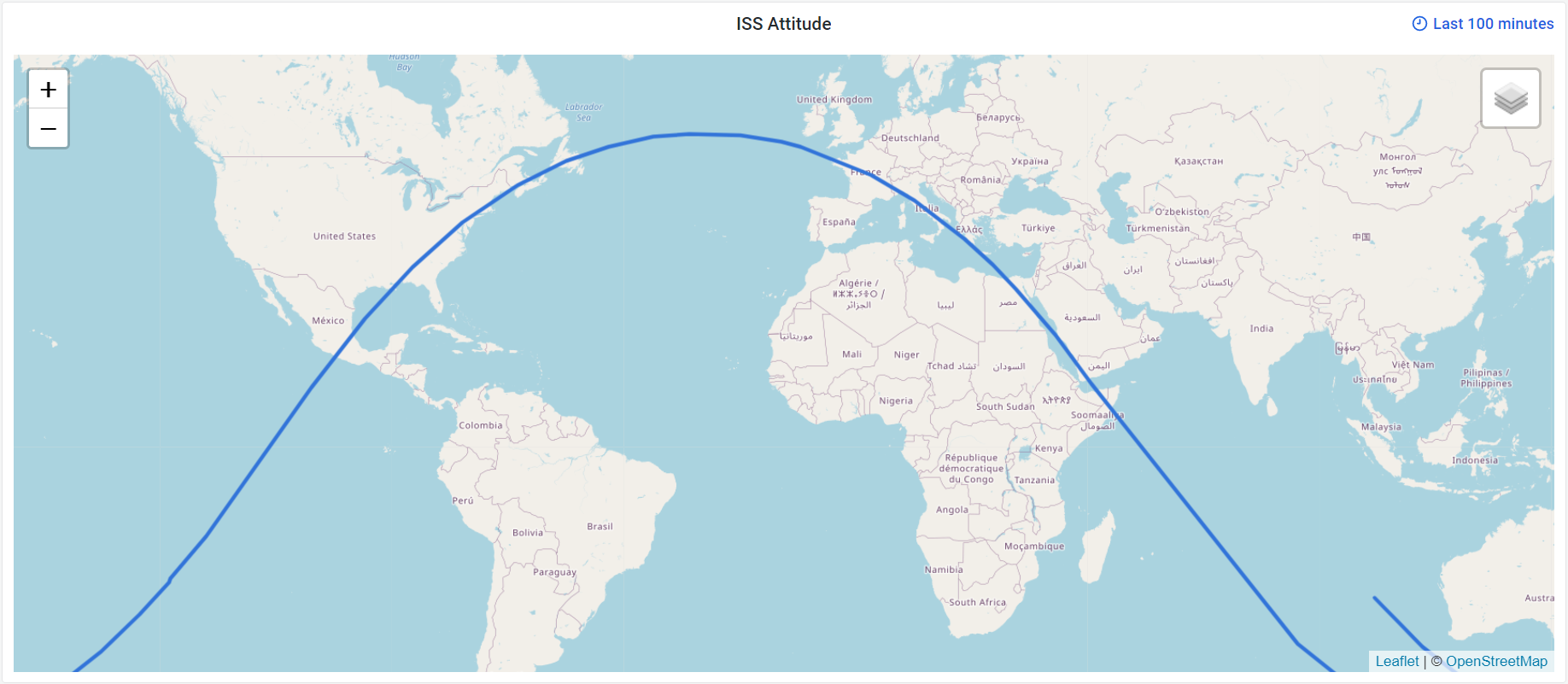}}
    \caption{An example of a Grafana panel being used to display the the ISS's GPS coordinates as a line on an interactive map using the TrackMap plugin \cite{CareyMetcalfe2022TrackMap}.}
    \label{fig:iss_location}
\end{figure}

Figure \ref{fig:homepage} shows the homepage of the new AMI, where plots are organized into dashboards, which are then further organized into panels on the homepage. Each panel is dedicated to a particular shift. This homepage can also be accessed remotely, facilitating remote shift taking. Grafana's wide range of visualization tools were used to replicate many of the original monitoring consoles at POCC, with several plugins being used to further extend the available tools. This allowed the creation of status maps (Figure \ref{fig:trd_occupancy}), histograms (Figure \ref{fig:tracker_calibrations}), scatter plots (Figure \ref{fig:L0_Channels}), time-series status indicators (Figure \ref{fig:pump_status}), table status indicators (Figure \ref{fig:tof_housekeeping}), on/off indicators (Figure \ref{fig:uttps_heater}), and GPS location markers (Figure \ref{fig:iss_location}).

\FloatBarrier

\FloatBarrier

\FloatBarrier

\FloatBarrier

\FloatBarrier

\FloatBarrier

\FloatBarrier

\FloatBarrier

\subsection{Scalability}
With the more modern method of having the server manage the database data streaming and client CPU manage the visualization aspects, the new AMI is configured for easy scalability. Previously, the legacy AMI managed both the database and visualization aspects, limiting the number of users at any given time due to CPU overload via visualization requests. With the visualizations offloaded to client CPUs, the number of people that can access the AMI with reasonably fast speeds concurrently has been significantly increased.

In addition, using the dual redundant backend setup, a load balancer can be applied to divert client requests to whichever backend server has the least load. While the current load levels do not require this, it can be enabled with relative ease, future proofing the new AMI for the growing number of users.

\subsection{Simplification of Data Retrieval for Analysis}
\label{section:simple_data}
Since Grafana can be used to directly retrieve a CSV file of data points, data analysis for certain measurements has been simplified, preventing the need for accessing AMS raw files directly for monitoring data.

A recent example of the new AMS Monitoring Interface being used in such a way was in the scheduling of the photon trigger. Since photons are neutral, the photon trigger activates when pair production occurs and a photon is converted into an electron and positron pair. However, their particle tracks are not isolated and often contaminated with background noise and particles, causing the photon trigger to fire too frequently and suppressing the activation of other triggers for other particles. To alleviate this issue, a schedule was necessary to understand when the chance of photons hitting the subdetectors was highest to enable the photon trigger only during those times. In order to do this, longitude, latitude, Level 1 Trigger times, and AMS livetimes was taken directly from the the new AMI to be processed. The result was a more sophisticated schedule of when to let the photon trigger be active, improving the quality of data acquisition. 

Another area of benefit is the easy applicability of machine learning algorithms. Plans to implement robotic monitoring systems that use machine learning to analyze the time series data and make predictions for the automatic adjustments are currently underway. Long short-term memory \cite{malhotra2015long} and Recurrent Neural Networks \cite{Su2019RobustRecurrent} specialize in data points with a temporal order to detect anomalous behavior, which can be used to trigger more sophisticated alerts automatically. The easy access to monitoring and science data via CSV files will help in creating training and testing datasets for these machine learning models.

\section{Conclusion}
The new AMS Monitoring Interface uses modern tools and design philosophies to better equip experts with the tools they need to monitor the AMS. It considers the AMS Raw Data points as time series data points, each with a timestamp, a measurement, and tags to sort and organize various sensor information. It uses InfluxDB as the backend database on a remote server to store, process, retrieve, and display all health data, and uses Grafana as the time series data analytics software to display interactive, dynamic plots on the client’s computer using the client's processing power. This significantly speeds up the displaying of plots, allows easy access to the AMI from across the globe, and simplifies the process of adding new measurements, future-proofing the AMI.

In addition, the new AMI allows users to download data points directly from the database via Grafana, eliminating the need for processing large AMS files. Grafana allows for the modification and plot building to be done graphically, eliminating the complexity of scripts and allowing different subdetector teams to be in charge of their own plot designs. Modern security features were implemented for secure access to AMS members around the globe. And lastly, because both InfluxDB and Grafana have APIs that can be used by simple scripts to automate tasks, two AMS Monitoring Interfaces were set up on AMS and CERN’s servers for dual redundancy and simple scripts were used to keep both instances of the AMI synchronized.

With a dedicated team formed that is responsible for the general upkeep and software updating of the new AMI, the new AMI was successfully deployed in January 2020 and has been in use since then at both the CERN and ASIA POCC.

\section*{Acknowledgements}
The authors would like to thank Professor Samuel C. C. Ting for his support, Dr. Michael Capell for his invaluable comments and review of the paper, and the AMS collaboration as a whole for their input and feedback during the design process of the new AMS Monitoring Interface. This research was supported by the Turkish Energy, Nuclear and Mineral Research Agency (TENMAK) under Grant No. 2020TAEK(CERN)A5.H1.F5-26.

\printcredits

\bibliographystyle{model1-num-names}

\bibliography{AMI_references}





\end{document}